\documentclass[aip,jcp,reprint,twocolumn,superscriptaddress,showpacs,floatfix]{revtex4-1}

\usepackage{mathrsfs}
\usepackage{amssymb}
\usepackage{amsmath}
\usepackage{dcolumn}
\usepackage{bm}
\usepackage{gensymb}
\usepackage{physics}
\usepackage{float}
\usepackage{multirow}
\usepackage{color}
\usepackage{tabularx}
\usepackage{hyperref}
\usepackage{graphicx}
\usepackage{amsmath}
\usepackage{graphicx}
\usepackage{dcolumn}
\usepackage{bm}
\usepackage{verbatim}
\usepackage[utf8]{inputenc}
\usepackage[T1]{fontenc}
\usepackage{mathptmx}
\usepackage{etoolbox}
\usepackage{subcaption} 
\usepackage{float}
\usepackage{bm}
\usepackage{mathrsfs}
\usepackage{multirow}
\usepackage{ragged2e}
\usepackage{hyperref}
\hypersetup{
    colorlinks=true,    
    linkcolor=blue,     
    citecolor=magenta,      
    pdfborder={0 0 0}   
}
\newcommand{\Cltwo}{Cl$_{2}$}
\newcommand{\Brtwo}{Br$_{2}$}
\newcommand{\Cltwoplus}{$\mathrm{(Cl_{2})}^{2+}$}
\newcommand{\Brtwoplus}{$\mathrm{(Br_{2})}^{2+}$}
\newcommand{\HBrtwoplus}{$\mathrm{(HBr)}^{2+}$}
\newcommand{\HItwoplus}{$\mathrm{(HI)}^{2+}$}

\begin{document}

\title{Reduced-Cost Four-Component Relativistic Double Ionization Potential Equation-of-Motion 
Coupled-Cluster Approaches with 4-Hole--2-Particle Excitations
and Three-Body Clusters}

\author{Tamoghna Mukhopadhyay}
\affiliation{Department of Chemistry, 
Indian Institute of Technology Bombay, Powai, Mumbai 400076, India}
\author{Madhubani Mukherjee}
\affiliation{Department of Chemistry, 
Indian Institute of Technology Bombay, Powai, Mumbai 400076, India}
\author{Karthik Gururangan}
\affiliation{Department of Chemistry,
Michigan State University, East Lansing, MI 48824, USA}
\author{Piotr Piecuch}
\email[e-mail: ]{piecuch@chemistry.msu.edu}
\affiliation{Department of Chemistry,
Michigan State University, East Lansing, MI 48824, USA}
\affiliation{Department of Physics and Astronomy,
Michigan State University, East Lansing, MI 48824, USA}
\author{Achintya Kumar Dutta}
\thanks{Corresponding author}
\email[e-mail: ]{achintya@chem.iitb.ac.in}
\affiliation{Department of Chemistry, 
Indian Institute of Technology Bombay, Powai, Mumbai 400076, India}
\affiliation{Department of Inorganic Chemistry, Faculty of Natural Sciences, Comenius University Bratislava
Ilkovičova 6, Mlynská dolina 842 15 Bratislava, Slovakia
}%
\begin{abstract}
The double ionization potential (DIP) equation-of-motion (EOM)
coupled-cluster (CC) method with 4-hole--2-particle 
(4$h$-2$p$) excitations on top of the CC with singles, 
doubles, and triples calculation, abbreviated as DIP-EOMCCSDT(4$h$-2$p$), along with its perturbative
DIP-EOMCCSD(T)(a)(4$h$-2$p$) approximation, are extended to
a relativistic four-component (4c) framework. 
In addition, we introduce and
test a new computationally practical
DIP-EOMCC approach, which we call
DIP-EOMCCSD(T)($\tilde{a}$)(4$h$-2$p$), that approximates the
treatment of 4$h$-2$p$ correlations within the DIP-EOMCCSD(T)(a)
(4$h$-2$p$) method and reduces the
$\mathscr{N}^8$ scaling
characterizing DIP-EOMCCSDT(4$h$-2$p$) and DIP-EOMCCSD(T)(a)(4$h$-2$p$) 
to $\mathscr{N}^7$ with the system size $\mathscr{N}$.
Further improvements in computational efficiency are obtained 
using the frozen natural spinor (FNS) approximation to reduce the 
numbers of unoccupied spinors entering the correlated steps of the 
DIP-EOMCC calculations according
to a well-defined occupation-number-based threshold.
The resulting 4c-FNS-DIP-EOMCC approaches are used to 
compute DIPs for the series of inert gas atoms from argon to radon
as well as the 
vertical DIPs in \Cltwo{}, \Brtwo{}, HBr, and HI, which
have been experimentally examined in the past. We demonstrate that,
when using complete basis set extrapolations and FNS truncation
thresholds of $10^{-4.5}$, the 
4c-FNS-DIP-EOMCCSD(T)($\tilde{a}$)(4$h$-2$p$) calculations are capable
of predicting DIPs in agreement with experimental data, 
improving upon their nonrelativistic and spin-free 
scalar-relativistic counterparts, particularly when examining DIPs characterized by stronger spin-orbit coupling effects.
\end{abstract}
\maketitle
\section{Introduction}
\label{sec1}
The accurate treatment of relativistic effects in chemical
systems has become an increasingly important facet of modern
computational chemistry. One application
of relativistic quantum chemical methods is the prediction 
of double ionization potentials (DIPs), which are critical to understanding
photoelectron and Auger electron spectroscopies.\cite{}
Indeed, spin-orbit coupling and other relativistic contributions can 
significantly impact core and valence
double ionization spectra 
in systems containing heavier atoms.\cite{alotibiEnhancedSpinOrbit2021,liEffectSpinorbitCoupling2025}
The accurate prediction of DIPs in such systems
remains a challenging
problem for many electronic structure methods due to the need to
treat relativistic interactions while capturing and balancing the many-electron correlation
effects characterizing the $N$- and ($N-2$)-electron species.
\cite{wladyslawskiLowLyingPotentialEnergy2002,nooijenStateSelectiveEquation2002,musialMultireferenceCoupledclusterTheory2011,kusUsingChargestabilizationTechnique2011,kusDeperturbativeCorrectionsChargestabilized2012,shenDoublyElectronattachedDoubly2013,shenDoublyElectronattachedDoubly2014}
While scalar-relativistic methods such as the zero-order regular 
approximation (ZORA),\cite{changRegularTwoComponentPauliLike1986,heullyDiagonalisationDiracHamiltonian1986,lentheRelativisticRegularTwocomponent1993} Douglas--Kroll--Hess (DKH) transformations,\cite{hessRelativisticElectronicstructureCalculations1986,jansenRevisionDouglasKrollTransformation1989,wolfGeneralizedDouglasKroll2002} and spin-free exact two-component (SFX2C) frameworks\cite{dyallInterfacingRelativisticNonrelativistic2001,liuExactTwocomponentHamiltonians2009} are widely used, more complete four-component (4c) approaches are preferable and can
serve as high-quality references, especially when dealing
with stronger relativistic effects.

Unfortunately, the application of 4c electron correlation methods
for describing double ionization 
suffers from increased computational costs due to
the use of spinor bases,
complex-valued Hamiltonians and wave functions, 
and the lack of spin
$S^2$ and $S_{z}$ symmetries. The use of uncontracted 
(e.g., Dyall-type \cite{dyallRelativisticDoublezetaTriplezeta2016,dyallRelativisticQuadrupleZetaRevised2006})
basis sets further exacerbates this issue.
In this work, we address this challenge by
developing \emph{ab initio} 4c approaches capable of
obtaining accurate DIPs in a computationally practical
fashion. 

Among the various techniques for computing DIPs in many-electron systems, the equation of motion (EOM) coupled-cluster (CC) method offers
an excellent balance between accuracy and computational cost, allowing one
to recover the exact, full configuration interaction (CI) results using a systematically improvable
hierarchy of approximations that can be performed using polynomial 
computational steps. As recently demonstrated in Ref.\ \onlinecite{dipeomccsdt}, 
the inclusion of one-, two-, and three-body clusters along with 2-hole (2$h$), 
3-hole--1-particle (3$h$-1$p$), and 4-hole--2-particle (4$h$-2$p$)
excitations in the DIP-EOMCC calculations, 
corresponding to the approach abbreviated as DIP-EOMCCSDT(4$h$-2$p$),
provides a highly accurate description of DIPs due to, in large part, 
achieving a well-balanced treatment of the correlation effects
characterizing the $N$- and $(N-2)$-electron states. In order to reduce the costs of the high-level 
DIP-EOMCCSDT(4$h$-2$p$) calculations, which scale as $\mathscr{N}^8$
with the system size $\mathscr{N}$, Ref.\ \onlinecite{dipeomccsdt} also introduced the 
perturbative DIP-EOMCCSD(T)(a)(4$h$-2$p$) approximation, which avoids the
expensive CC calculation with singles, doubles, and triples (CCSDT),
\cite{ccsdt1986,nogaFullCCSDTModel1987,scuseriaNewImplementationFull1988,wattsCoupledclusterSingleDouble1990}
used to describe the $N$-electron ground-state, by accounting for the 
effects of three-body clusters using perturbative arguments inspired by
the CCSD(T)(a)-based approach of Ref.\ \onlinecite{eomccsdta}.
It was demonstrated that the DIP-EOMCCSD(T)(a)(4$h$-2$p$)
method accurately reproduces the DIPs obtained using its
DIP-EOMCCSDT(4$h$-2$p$) parent for
several diatomic molecules near their equilibrium geometries.

Given the computational challenges associated
with the 4c framework, the adoption of high-level fully
relativistic DIP-EOMCC methodologies has been comparatively
slower. The previous 4c DIP-EOMCC approaches
of Refs.\ \onlinecite{Pathak2014,pathakRelativisticDoubleionizationEquationofmotion2020,wangEquationofmotionCoupledclusterMethod2015} were limited
to the DIP-EOMCCSD(3$h$-1$p$) level,
\cite{wladyslawskiLowLyingPotentialEnergy2002,nooijenStateSelectiveEquation2002,musialMultireferenceCoupledclusterTheory2011,kusUsingChargestabilizationTechnique2011,kusDeperturbativeCorrectionsChargestabilized2012,%
shenDoublyElectronattachedDoubly2013,shenDoublyElectronattachedDoubly2014}
which treats
2$h$ and 3$h$-1$p$ excitations on top of CC with singles
and doubles (CCSD).
\cite{purvisFullCoupledclusterSingles1982,cullenLinkedSinglesDoubles1982,scuseriaClosedshellCoupledCluster1987,piecuchOrthogonallySpinadaptedCoupledcluster1989}
More recently, the high-level DIP-EOMCCSDT(4$h$-2$p$) approach
was extended to the two-component relativistic regime in
Ref.\ \onlinecite{liRelativisticTwocomponentDouble2025}, however, due to
excessive CPU and memory requirements, the resulting DIP-EOMCCSDT(4$h$-2$p$)
calculations were limited to triple-zeta-quality basis sets.
Thus, a major aim of the present study is to develop
4c DIP-EOMCC approaches that incorporate up to 4$h$-2$p$ correlations
and three-body clusters
in a robust and practical fashion, and which can be
applied to systems described with larger quadruple-zeta (QZ) 
basis sets.
To accomplish this task, we follow two strategies 
for reducing computational costs. First, we simplify the treatment 
of 4$h$-2$p$ excitations
in the DIP-EOMCCSD(T)(a)(4$h$-2$p$) method to obtain a new approach
abbreviated as DIP-EOMCCSD(T)($\tilde{a}$)(4$h$-2$p$), which can be performed
using computational steps that scale as $\mathscr{N}^7$ with
the system size. In addition, we adopt 
the frozen natural spinor (FNS)
technique,\cite{chamoliReducedCostFourcomponent2022,yuanAssessingMP2Frozen2022}
as implemented in Ref. \onlinecite{4c-eomcc-fns},
to reduce the numbers of virtual spinors entering the 4c-DIP-EOMCC
calculations in a controlled and systematic fashion.
The resulting FNS-based 4c-DIP-EOMCCSD(T)($\tilde{a}$)(4$h$-2$p$) calculations
are applied to obtain the DIPs of the series of inert gas atoms from argon to radon as well as the vertical DIPs in 
\Cltwo{}, \Brtwo{}, HBr, and HI using up to uncontracted
Dyall-type QZ-quality basis sets containing up to 
764 
orbitals.
This allows us to carry out complete basis set (CBS)
extrapolations
in order to make more meaningful comparisons with the existing
experimental data.
\section{Theory}
\label{sec2}
\subsection{The Treatment of Relativistic Effects}
\label{sec2.1}
Relativistic interactions in quantum chemistry are typically described using a 4c Dirac--Coulomb (DC) Hamiltonian, which
for an $N$-electron system with $M$ clamped nuclei is given by (assuming atomic units)
\begin{equation}
\label{eq1}
\hat{H}^{\text{DC}} = \sum_{i=1}^{N} \left[ c \vec{\alpha}_i \cdot \vec{p}_i + \beta_{i} m_{0} c^2 + \sum_{A=1}^{M} \hat{V}_{iA} \right] + \sum_{j>i = 1}^{N} \dfrac{1}{r_{ij}} \hat{I}_{4}, 
\end{equation}
where $c$ is the speed of light, $m_{0}$ is the rest mass of an electron, and $\hat{V}_{iA}$ is the electrostatic attraction
between electron $i$ and nucleus $A$. As usual, $\vec{p}_i$ is the momentum of the $i$th electron
and $\vec{\alpha}_{i}$ and $\beta_{i}$ denote the Dirac matrices associated with electron $i$. 
The operator $\hat{I}_{4}$ in Eq.\ (\ref{eq1}) is a 4$\times$4 identity matrix.
A zeroth-order description of the many-electron system is obtained by solving the Diract--Hartree--Fock (DHF)
mean-field equations. The matrix DHF equations are expressed as 
\begin{equation}
\label{eq2}
\begin{pmatrix}
	\hat{V} + \hat{J} - \hat{K} & c(\boldsymbol{\sigma} \cdot \hat{p}) - \hat{K} \\
	c(\boldsymbol{\sigma} \cdot \hat{p}) - \hat{K} & \hat{V} - 2m_{0}c^2 + \hat{J} - \hat{K} 
\end{pmatrix}
\begin{pmatrix}
	\phi^{L} \\ \phi^{S}
\end{pmatrix}
= E
\begin{pmatrix}
	\phi^{L} \\ \phi^{S}
\end{pmatrix},
\end{equation}
where $\phi^{L}$ ($\phi^{S}$) refers to the large (small) component of the 4c spinor $\psi$. The operators $\hat{V}$, $\hat{J}$,
and $\hat{K}$ in Eq.\ (\ref{eq2}) denote the electron-nuclear, Coulomb, and exchange potentials, respectively. While the DC
Hamiltonian serves as the natural starting point for relativistic quantum chemical calculations, one can also consider the
Gaunt (DCG) and Breit (DCB) corrections to the DC Hamiltonian, which are given by
\begin{equation}
\label{eq1.1}
{{\hat{H}}^{\text{DCG}}}=\sum\limits_{i=1}^{N}{\left[ c{{{\vec{\alpha }}}_{i}}.{{{\vec{p}}}_{i}}+{{\beta }_{i}}{{m}_{0}}{{c}^{2}}+\sum\limits_{A=1}^{M}{{{{\hat{V}}}_{iA}}}\right]}+\sum\limits_{j>i=1}^{N}({\frac{1}{{{r}_{ij}}}+G_{ij}){{{\hat{I}}}_{4}}}
\end{equation}
and
 \begin{equation}
\label{eq1.1}
 {{\hat{H}}^{\text{DCB}}}=\sum\limits_{i=1}^{N}{\left[ c{{{\vec{\alpha }}}_{i}}.{{{\vec{p}}}_{i}}+{{\beta }_{i}}{{m}_{0}}{{c}^{2}}+\sum\limits_{A=1}^{{M}}{{{{\hat{V}}}_{iA}}}\right]}+\sum\limits_{j>i=1}^{N}({\frac{1}{{{r}_{ij}}}+B_{ij}){{{\hat{I}}}_{4}}},
\end{equation}
respectively, where 
\begin{equation}
    G_{ij} = -\frac{\alpha_i \cdot \alpha_j}{2 r_{ij}}
\end{equation}
and 
\begin{equation}
    B_{ij} = -\frac{1}{2 r_{ij}} 
\left[ \alpha_i \cdot \alpha_j 
+ \frac{(\alpha_i \times r_{ij}) \cdot (\alpha_j \times r_{ij})}{r_{ij}^2} \right].
\end{equation}

After obtaining the DHF mean-field state via Eq.\ (\ref{eq2}) using either the DC, DCG, or DCB Hamiltonian, the 
missing many-electron correlation effects can be incorporated using the
no-pair approximation.\cite{sucherFoundationsRelativisticTheory1980}
We note that when using the DCG or DCB Hamiltonians, the Gaunt or Breit corrections 
are only included in the DHF step, and are neglected in the subsequent integral
transformation.
\subsection{Overview of the Double Ionization Potential Equation-of-Motion Coupled-Cluster
Method}
\label{sec2.2}
In order to treat many-electron correlation
effects on top of the DHF mean-field solution,
we rely on the hierarchy of approximations
based on the CC theory alongside its
DIP-EOMCC extension to doubly ionized
states.
In the single-reference CC theory,
\cite{coesterBoundStatesManyparticle1958,coesterShortrangeCorrelationsNuclear1960,cizekCorrelationProblemAtomic1966,cizekUseClusterExpansion1969,paldusCorrelationProblemsAtomic1972}
the ground-state many-body wave function for an $N$-electron
system is described using the
exponential wave function ansatz \cite{Hubbard:1957,Hugenholtz:1957}
\begin{equation}
\label{eq3}
|\Psi_{0}^{(N)}\rangle =e^{\hat{T}}|\Phi\rangle,
\end{equation}
where $|\Phi\rangle$ is the DHF determinant, which serves as a
Fermi vacuum, and $\hat{T}$ is the cluster operator,
\begin{equation}
\label{eq4}
    \hat{T} = \sum_{n=1}^{M_{T}} \hat{T}_{n},
\end{equation}
where the $n$-body component of $\hat{T}$ is
\begin{equation}
\label{eq5}
\hat{T}_{n} = \sum_{\substack{i_{1}<\cdots<i_{n} \\ a_{1}<\cdots<a_{n}}} t_{a_{1}\ldots a_{n}}^{i_{1}\ldots i_{n}} \hat{a}^{a_{1}}\ldots \hat{a}^{a_{n}} \hat{a}_{i_{n}}\ldots \hat{a}_{i_{1}}.
\end{equation}
As usual, indices $i_1, i_2, \ldots$ ($a_1, a_2, \ldots$) denote the 
spinors that are occupied (unoccupied) in $|\Phi\rangle$ and $\hat{a}^{p}$ ($\hat{a}_{p}$) 
represents the fermionic creation (annihilation) operator 
associated with the spinor $|p\rangle$. 
The value $M_{T}$ in Eq.\ (\ref{eq4}) controls the truncation in $\hat{T}$, which gives rise to the conventional hierarchy of CC
approximations. For example, 
$M_{T}=2$ defines the basic CCSD method, in which
$\hat{T} = \hat{T}_{1} + \hat{T}_{2}$, while $M_{T}=3$ yields the higher-level
CCSDT approach with $\hat{T} = \hat{T}_{1} + \hat{T}_{2} + \hat{T}_{3}$, and so on.
For a given truncation $M_{T}$, the amplitudes characterizing
the cluster operator $T$ are determined by solving the
projective conditions,
\begin{equation}
\label{eq8}
\langle \Phi_{i_{1}\ldots i_{n}}^{a_{1}\ldots a_{n}} | \bar{H}|\Phi \rangle = 0,\:\:i_{1}<\ldots i_{n},\:\:a_{1}<\ldots<a_{n},
\end{equation}
for $n=1,\ldots,M_{T}$,
where
\begin{equation}
\label{eq9}
    \bar{H} = e^{-\hat{T}} \hat{H} e^{\hat{T}}
\end{equation}
is the corresponding similarity-transformed Hamiltonian. After
solving Eq.\ (\ref{eq8}) to obtain the cluster amplitudes
$t_{a_{1}\ldots a_{n}}^{i_{1}\ldots i_{n}}$, $n = 1,\ldots,M_{T}$, the 
ground-state energy is obtained \emph{a posteriori} as
\begin{equation}
\label{eq7}
    E_{0} = \langle  \Phi | \bar{H} | \Phi \rangle.
\end{equation}

It has been well established over the 
course of many studies
employing 
nonrelativistic (NR) Hamiltonians
(cf. Refs.\ 
\onlinecite{paldus-li,bartlett-musial2007} for representative
examples)
that 
the
higher-level CC approximations, like CCSDT ($M_{T}=3$) and CCSDTQ ($M_{T}=4$),
\cite{oliphantCoupledclusterMethodTruncated1991,kucharskiRecursiveIntermediateFactorization1991,kucharskiCoupledclusterSingleDouble1992,piecuchStateselectiveMultireferenceCoupledcluster1994}
are capable of recovering a highly accurate treatment of the many-electron
correlation effects relative to the exact, full CI,
solution in most
chemically relevant problems,
including noncovalent interactions, bond dissociations, and open shells like radicals and biradicals.
The same is true when examining the 
convergence of the CC hierarchy applied to relativistic
Hamiltonians.\cite{porsevTripleExcitationsRelativistic2006,fabbroHighlyAccurateExpectation2025}

In order to describe double ionization within the CC framework, 
we can turn to the DIP-EOMCC methodology, in which the ground ($\mu=0$)
and excited ($\mu>0$) states of the $(N-2)$-electron
target system are described as 
\begin{equation}
    |\Psi_{\mu}^{(N-2)}\rangle = \hat{R}_{\mu}^{(-2)} |\Psi_{0}^{(N)}\rangle,
\end{equation}
where the doubly ionizing operator
\begin{equation}
\label{eqRk}
    \hat{R}_{\mu} = \sum_{n=0}^{M_{R}} \hat{R}_{\mu,(n+2)h\mbox{-}np}
\end{equation}
consists of many-body components 
\begin{align}
\label{eqRkn}
    \hat{R}_{\mu,(n+2)h\mbox{-}np} = &\sum_{\substack{i<j<k_1<\ldots<k_n\\c_{1}<\ldots<c_{n}}} 
    r_{\phantom{ab}c_1\cdots c_n}^{ijk_{1}\cdots k_{n}}(\mu) \\ &\times
    \hat{a}^{c_{1}} \ldots \hat{a}^{c_{n}}\hat{a}_{k_{n}} \ldots  \hat{a}_{j} \hat{a}_{i}
\end{align}
that remove two electrons from the $N$-electron ground-state wave
function $|\Psi_{0}^{(N)}\rangle$ via $(n+2)h$-$np$ excitations. The
truncation parameter $M_{R}$ in Eq.\ (\ref{eqRk}) determines the maximum
level of $(n+2)h$-$np$ excitations included in $\hat{R}_{\mu}^{(-2)}$. By
varying the values of $M_{T}$ and $M_{R}$, 
we obtain the standard
hierarchy of DIP-EOMCC approximations.

For example, DIP-EOMCC methods based on a
CCSD description of the $N$-electron system ($M_{T}=2$)
include the DIP-EOMCCSD(3$h$-1$p$) and DIP-EOMCCSD(4$h$-2$p$)
\cite{shenDoublyElectronattachedDoubly2013,shenDoublyElectronattachedDoubly2014}
approaches, which include up to the 3$h$-1$p$ ($M_{R}=1$) and
4$h$-2$p$ ($M_{R}=2$) excitations in the ($N-2$)-electron species, respectively.
One can also consider DIP-EOMCC methods based on the more accurate
$N$-electron CCSDT state ($M_{T}=3$), such as the DI-EOMCCSDT 
scheme\cite{musialMultireferenceCoupledclusterTheory2011}, which treats
the 2$h$ and 3$h$-1$p$ excitations ($M_{R}=1$) on top of CCSDT,
and the recently introduced DIP-EOMCCSDT(4$h$-2$p$) approach corresponding
to $M_{T}=3$ and $M_{R}=2$ that provides a full treatment 
of both 4$h$-2$p$ and $\hat{T}_{3}$ correlations.
While all of the aforementioned DIP-EOMCC methods are convenient 
tools for determining DIPs in many-electron systems, the study in 
Ref.\ \onlinecite{dipeomccsdt} emphasizes that obtaining highly accurate DIPs
requires that one balances the correlations due to $(n+2)h$-$np$ excitations
in the $(N-2)$-electron target states with the CC treatment of the 
underlying $N$-electron system. 
In particular, when 4$h$-2$p$ excitations are included in the $\hat{R}_{\mu}^{(-2)}$ operator,
which is often necessary for obtaining accurate energetics,
\cite{shenDoublyElectronattachedDoubly2013,shenDoublyElectronattachedDoubly2014,dipeomccsdt}
one must also account for $\hat{T}_{3}$ correlations in the $N$-electron ground state.

A standard DIP-EOMCC calculation consists of solving the
matrix eigenvalue problem, which for $M_{R} \leq M_{T}$ (a condition
required to
obtain size-intensive DIPs\cite{shenDoublyElectronattachedDoubly2013,shenDoublyElectronattachedDoubly2014,jspp-dea-dip-2021}), is given by
\begin{equation}
\label{eomeig}
[\bar{H}_{\text{open}}, \hat{R}_{\mu}^{(-2)}]|\Phi\rangle = \omega_{\mu}^{(N-2)} \hat{R}_{\mu}^{(-2)}|\Phi\rangle,
\end{equation}
where $\bar{H}_{\text{open}}$ refers to the diagrams in $\bar{H}$ that contain
external fermion lines. The eigenvalues $\omega_{\mu}^{(N-2)}$ obtained
by solving Eq.\ (\ref{eomeig}) are the DIPs corresponding to vertical
transitions between the $N$-electron ground state $|\Psi_{0}^{(N)}\rangle$
and the ground ($\mu=0$) and excited ($\mu>0$) states of the $(N-2)$-electron
system, $|\Psi_{\mu}^{(N-2)}\rangle$, while the corresponding right-eigenvectors
provide the excitation amplitudes $r_{\phantom{ab}c_1\cdots c_n}^{ijk_{1}\cdots k_{n}}(\mu)$,
for $n=0,\ldots,M_{R}$, characterizing the $\hat{R}_{\mu}^{(-2)}$ operator.
Thus, the post-DHF steps of a 4c DIP-EOMCC calculation consist of solving 
Eq.\ (\ref{eq8}) to obtain the truncated form of the cluster operator
$\hat{T}$ and energy $E_{0}$ [Eq.\ (\ref{eq7})]
characterizing the $N$-electron ground state $|\Psi_{0}^{(N)}\rangle$
and diagonalizing the corresponding similarity-transformed
Hamiltonian $\bar{H}$ [Eq.\ (\ref{eq9})] in the appropriate ($N-2$)-electron
subspace of the Fock space associated with the content of $\hat{R}_{\mu}^{(-2)}$ 
following Eq.\ (\ref{eomeig}).

In the full DIP-EOMCCSDT(4$h$-2$p$) method, which is the highest level
of DIP-EOMCC theory considered in this work, the cluster operator is
given by $\hat{T} = \hat{T}_{1} + \hat{T}_{2} + \hat{T}_{3}$ and the doubly ionizing operator
is truncated at the 4$h$-2$p$ level to yield 
$\hat{R}_{\mu}^{(-2)} = \hat{R}_{\mu,2h} + \hat{R}_{\mu,3h\mbox{-}1p} + \hat{R}_{\mu,4h\mbox{-}2p}$. 
As a result, the DIP-EOMCCSDT(4$h$-2$p$) calculation involves computational steps
that scale as $n_{o}^3 n_{u}^5$, corresponding to the preliminary CCSDT
calculation for the underlying $N$-electron system, followed by the
diagonalization of the Hamiltonian in the subspace of the Fock space
associated with $\hat{R}_{\mu}^{(-2)}$, which scale as $n_{o}^4 n_{u}^4$, where 
$n_{o}$ ($n_{u}$) denotes the number of occupied (unoccupied) spinors
in $|\Phi\rangle$. Both of these steps scale as $\mathscr{N}^8$ with
the system size $\mathscr{N}$, however, the preliminary CCSDT step is
significantly more expensive due to the $n_{u}^5$ scaling with the
number of unoccupied spinors. 

The DIP-EOMCCSD(T)(a)(4$h$-2$p$) approximation to DIP-EOMCCSDT(4$h$-2$p$)
introduced in Ref.\ \onlinecite{dipeomccsdt} provides on solution to this
issue by replacing the CCSDT step by its much less expensive CCSD analog, which
involves computational steps that scale as $n_{o}^2 n_{u}^4$. In DIP-EOMCCSD(T)(a)(4$h$-2$p$),
the $T_{3}$ cluster is approximated according to perturbation theory,
\begin{equation}
    \hat{T}^{[2]}_3 = \hat{D}_3[\hat{V}_{N}, \hat{T}_2],
    \label{eqt3pert}
\end{equation}
where $\hat{T}_{2}$ in Eq.\ (\ref{eqt3pert}) refers to the two-body component of $\hat{T}$ 
obtained from the CCSD calculation, and we have adopted the 
M{\o}ller--Plesset (MP) partitioning of the electronic Hamiltonian, 
$\hat{H}_{N} = \hat{H} - \langle \Phi | \hat{H} | \Phi \rangle = \hat{F}_{N} + \hat{V}_{N}$, 
with $\hat{F}_{N}$ and $\hat{V}_{N}$ representing the usual Fock and fluctuation operators 
resulting from normal ordering of the Hamiltonian 
with respect to $|\Phi\rangle$, and $\hat{D}_3$ is the three-body MP energy denominator.
The one- and two-body components of $\hat{T}$ determined in the $N$-electron CCSD calculation 
are then noniteratively corrected for the $\hat{T}_{3}$ effects via
\begin{equation}
    \hat{T}^{\prime}_1 =  \hat{T}_1 + \hat{D}_1[\hat{V}_{N}, \hat{T}^{[2]}_3]
    \label{eqt1corr}
\end{equation}
and
\begin{equation}
    \hat{T}^{\prime}_2 =  \hat{T}_2 + \hat{D}_2[ \hat{H}_{N}, \hat{T}^{[2]}_3],
    \label{eqt2corr}
\end{equation}
where $\hat{D}_{1}$ and $\hat{D}_{2}$ denote the 1- and 2-body MP energy denominators. 
Using $\hat{T}_{1}^{\prime}$, $\hat{T}_{2}^{\prime}$, and $\hat{T}_{3}^{[2]}$, the 
CCSD(T)(a) similarity-transformed Hamiltonian,\cite{eomccsdta}
represented in the ($N-2$)-electron subspace of the Fock space spanned by
the 2$h$, 3$h$-1$p$, and 4$h$-2$p$ excitations,
\begin{widetext}
\begin{equation}
\label{eqHccsdta}
\bar{H}^{[\text{CCSD(T)(a)}]} = 
\begin{pmatrix}
\bar{H}^{\prime}_{2h,2h} & \bar{H}^{\prime}_{2h,3h\mbox{-}1p} & \bar{H}^{\prime}_{2h,4h\mbox{-}2p}\\
\bar{H}^{\prime}_{3h\mbox{-}1p,2h} + [\hat{V}_{N}, \hat{T}_{3}^{[2]}]& \bar{H}^{\prime}_{3h\mbox{-}1p,3h\mbox{-}1p} & \bar{H}^{\prime}_{3h\mbox{-}1p,4h\mbox{-}2p}\\
\bar{H}^{\prime}_{4h\mbox{-}2p,2h} + [\hat{F}_{N} + \hat{V}_{N}, \hat{T}_{3}^{[2]}] & \bar{H}^{\prime}_{4h\mbox{-}2p,3h\mbox{-}1p} + [\hat{V}_{N}, \hat{T}_{3}^{[2]}] & \bar{H}^{\prime}_{4h\mbox{-}2p,4h\mbox{-}2p} 
\end{pmatrix},
\end{equation}
\end{widetext}
where 
$\bar{H}^{\prime} = e^{-\hat{T}_{1}^{\prime}-\hat{T}_{2}^{\prime}} \hat{H} e^{\hat{T}_{1}^{\prime}+\hat{T}_{2}^{\prime}}$,
can be constructed and diagonalized to obtain the vertical DIPs and excitation amplitudes
characterizing the $\hat{R}_{\mu}^{(-2)}$ operator. In this way, 
the DIP-EOMCCSD(T)(a)(4$h$-2$p$) approximation avoids
the expensive $N$-electron CCSDT calculation and reduces the 
$n_{o}^3 n_{u}^5$ costs characterizing its DIP-EOMCCSDT(4$h$-2$p$) parent
to $n_{o}^4 n_{u}^4$.
As shown in Ref.\ \onlinecite{dipeomccsdt}, the DIPs obtained using
the DIP-EOMCCSD(T)(a)(4$h$-2$p$) approach are very close to those
computed with DIP-EOMCCSDT(4$h$-2$p$) when examining
the vertical transitions in closed-shell molecules near their
equilibrium ground-state structures.

In the 4c framework adopted in the present study,
the $n_{o}^4 n_{u}^4$ steps entering the 
DIP-EOMCCSD(T)(a)(4$h$-2$p$) calculations proved to be very 
computationally demanding, especially for larger QZ basis sets, 
so we invoked three additional approximations motivated by practicality, resulting in the
method abbreviated as DIP-EOMCCSD(T)($\tilde{a}$)(4$h$-2$p$). 
In DIP-EOMCCSD(T)($\tilde{a}$)(4$h$-2$p$), the CCSD(T)(a)
similarity-transformed Hamiltonian is further simplified 
by
(i) removing all contributions to 3-body components of $\bar{H}^{[\text{CCSD(T)(a)}]}$
arising from contractions with $\hat{T}_{3}$ clusters,
(ii) neglecting all $\hat{T}_{3}$ contributions
in the projections
corresponding onto 4$h$-2$p$ determinants,
and (iii) assuming that $\bar{H}^{[\text{CCSD(T)(a)}]}$
is quasi-diagonal in the 4$h$-2$p$ sector of the Fock space,
allowing us to replace $\bar{H}^{\prime}_{4h\mbox{-}2p,4h\mbox{-}2p}$ 
by its zeroth-order, MP-like, counterpart.
The programmable expressions for the DIP-EOMCCSD(T)($\tilde{a}$)(4$h$-2$p$) sigma equations corresponding to 
projections onto 2$h$ ($|\Phi_{ij}\rangle$), 
3$h$-1$p$ ($|\Phi_{ijk}^{\phantom{ab}c}\rangle$),
and 4$h$-2$p$ ($|\Phi_{ijkl}^{\phantom{ab}cd}\rangle$)
determinants, as implemented in the BAGH program package, are
\begin{widetext}
\begin{equation}
\langle \Phi_{ij} | (\overline{H}_{N,\mathrm{open}}^{(\text{CCSDT})} R_{\mu}^{(-2)})_C | \Phi \rangle 
= \mathscr{A}^{ij}[
-\bar{h}_{m}^{i} r_{}^{mj}(\mu)
+ \tfrac{1}{4} \bar{h}_{mn}^{ij}r_{}^{mn}(\mu) 
+ \tfrac{1}{2} \bar{h}_{m}^{e}r_{\phantom{ab}e}^{ijm}(\mu) 
- \tfrac{1}{2} \bar{h}_{mn}^{if}r_{\phantom{ab}f}^{mjn}(\mu)
+ \tfrac{1}{8} \bar{h}_{mn}^{ef}r_{\phantom{ab}ef}^{ijmn}(\mu)],
\label{eq2h}
\end{equation}
\begin{align}
\langle \Phi_{ijk}^{\phantom{ab}c} | (\overline{H}_{N,\mathrm{open}}^{(\text{CCSDT})} R_{\mu}^{(-2)})_C | \Phi \rangle 
=&\mathscr{A}^{ijk}[
\tfrac{1}{2} {I^{\prime}}^{ie}(\mu) t_{ec}^{jk} 
- \tfrac{1}{2} \bar{h}_{cm}^{ki} r_{}^{mj}(\mu)
+ \tfrac{1}{6} \bar{h}_{c}^{e} r_{\phantom{ab}e}^{ijk}(\mu)
- \tfrac{1}{2} \bar{h}_{m}^{k} r_{\phantom{ab}c}^{ijm}(\mu)
+ \tfrac{1}{4} \bar{h}_{mn}^{ij} r_{\phantom{ab}c}^{mnk}(\mu)
\nonumber \\
&+ \tfrac{1}{2} \bar{h}_{cm}^{ke} r_{\phantom{ab}e}^{ijm}(\mu)
+ \tfrac{1}{6} \bar{h}_{m}^{e} r_{\phantom{ab}ce}^{ijkm}(\mu)
- \tfrac{1}{4} \bar{h}_{mn}^{kf} r_{\phantom{ab}cf}^{ijmn}(\mu)
,
\label{eq3h1p}
\end{align}
and
\begin{align}
\langle \Phi_{ijkl}^{\phantom{ab}cd} | (\overline{H}_{N,\mathrm{open}}^{(\text{CCSDT})} R_{\mu}^{(-2)})_C | \Phi \rangle 
= &\mathscr{A}^{ijkl}\mathscr{A}_{cd}[
\tfrac{1}{12} \bar{h}_{dc}^{le} r_{\phantom{ab}e}^{ijk}(\mu)
- \tfrac{1}{4} \bar{h}_{dm}^{lk} r_{\phantom{ab}c}^{ijm}(\mu)
- \tfrac{1}{12} I_{\phantom{ab}m}^{ijk}(\mu) t_{cd}^{ml}
+ \tfrac{1}{4} I_{\phantom{ab}c}^{ije}(\mu) t_{ed}^{kl}
+ \tfrac{1}{24} f_{d}^{e} r_{\phantom{ab}ce}^{ijkl}(\mu)
\nonumber \\
&- \tfrac{1}{12} f_{m}^{i} r_{\phantom{ab}cd}^{mjkl}(\mu),
\label{eq4h2p}
\end{align}
\end{widetext}
where the index antisymmetrizers in Eq.\ (\ref{eq2h})--(\ref{eq4h2p}) are
$\mathscr{A}^{pq} = \mathscr{A}_{pq} = 1 - (pq)$, 
$\mathscr{A}^{pqr} = \mathscr{A}^{p/qr}\mathscr{A}^{qr}$, 
and $\mathscr{A}^{pqrs} = \mathscr{A}^{p/qrs}\mathscr{A}^{qrs}$, 
with the partial antisymmetrizers defined as 
$\mathscr{A}^{p/qr} = 1 - (pq) - (pr)$ and 
$\mathscr{A}^{p/qrs} = 1 - (pq) - (pr) - (ps)$. The quantities
$f_{p}^{q}$ in Eq.\ (\ref{eq4h2p}) denote the standard
one-electron Fock matrix, while the expressions
for the remaining
one-body ($\bar{h}_{p}^{q}$) and two-body ($\bar{h}_{pq}^{rs}$) 
components of the similarity-transformed Hamiltonian
as well as additional intermediates entering 
Eqs.\ (\ref{eq2h})--(\ref{eq4h2p})
are provided in the Supplementary material.
\subsection{The Frozen Natural Spinor Technique}
\label{sec2.3}
The natural spinors\cite{chamoliReducedCostFourcomponent2022} are the relativistic analogue of the NR natural orbitals, introduced by Löwdin.\cite{lowdinQuantumTheoryManyParticle1955} Natural spinors are obtained as the eigenfunctions of the relativistic correlated one-body reduced density matrix.\cite{chamoliRelativisticReducedDensity2024} Among the various schemes for
obtaining natural spinors,\cite{chamoliReducedCostFourcomponent2022,mukhopadhyayReducedcostRelativisticEquationofMotion2025,chakraborty2025low} we have chosen the standard MP2-based natural spinors,\cite{chamoliReducedCostFourcomponent2022} which is obtained by 
rotating the set of unoccupied DHF
spinors using the eigenvectors of the
virtual-virtual block of the 
one-body reduced density matrix 
computed at the MP2 level.

After constructing  the
MP2 one-body reduced density matrix, it is diagonalized and the eigenvalues are occupancies of the corresponding virtual natural spinors (eigenvectors).
\begin{equation}
\label{eq24}
    Dv=Vn
\end{equation}
Sorting the natural spinors based on their occupancies in a decreasing order leads to a gradual hierarchy of their contribution to the correlation. One can set up a predefined threshold ($n_{\text{thresh}}$) to truncate them, where only the natural spinors with occupancies larger than $n_{\text{thresh}}$ are considered and the rest of them are dropped off in the following calculations. Truncation
of the virtual space
can be accomplished by multiplying the natural spinor
transformation matrix $V$ by a thresholding matrix $\tau$
according to
\begin{equation}
\label{eq25}
    \tilde{V}=V\tau,
\end{equation}
where
\begin{align}
\label{eq26}
    & {\tau_{ij}}={{\delta }_{ij}}\quad &\forall {{n}_{i}}>{{n}_{thresh}} \\ 
    & {\tau_{ij}}=0\quad &\forall {{n}_{i}}\le {{n}_{thresh}}. \\ 
\end{align}

The virtual-virtual block of the Fock matrix is then transformed into the natural spinor basis according to
\begin{equation}
\label{eq27}
    \tilde{F}={{\tilde{V}}^{\dagger }}F\tilde{V}.
\end{equation}
Here, a tilde above an operator
refers to the operator expressed in
the truncated basis. Diagonalizing $\tilde{F}$ leads to natural spinor energies $\left(\tilde{\epsilon }\right)$ as eigenvalues and eigenvectors $\left(\tilde{Z}\right)$, which are used to semi-canonicalize the new basis,
\begin{equation}
\label{eq28}
\tilde{F}\tilde{Z}=\tilde{Z}\tilde{\epsilon}.
\end{equation}

In practice, the transformation
between the original virtual molecular 
spinors and the truncated virtual
natural spinor basis is
\begin{equation}
\label{eq29}
    \tilde{B}=\tilde{V}\tilde{Z}.
\end{equation}
The atomic spinor integrals can be directly converted to the truncated natural spinor basis by the following transformation matrices,
\begin{align}
\label{eqao2fns}
    & {{{\tilde{U}}}_{occ}}={{U}_{occ}} \\ 
    & {{{\tilde{U}}}_{vir}}={{U}_{vir}}\tilde{V}\tilde{Z}={{U}_{vir}}\tilde{B}, \\ \nonumber
\end{align}
where $\hat{U}$ represents the 
transformation matrix between the
atomic spinor basis and the
molecular spinor (DHF) basis. 
This approach is called FNS,\cite{chamoliReducedCostFourcomponent2022,yuanAssessingMP2Frozen2022,chamoliFrozenNaturalSpinors2025,majeeReducedCostFourcomponent2024,mukhopadhyayReducedcostRelativisticEquationofMotion2025} as the occupied sector is kept frozen at the DHF level of theory.

\section{Computational Details}
\label{sec3}
The 4c DIP-EOMCCSDT(4$h$-2$p$), DIP-EOMCCSDT(a)(4$h$-2$p$),
and DIP-EOMCCSD(T)($\tilde{a}$)(4$h$-2$p$) methods along with
their FNS-truncated counterparts are implemented in BAGH\cite{duttaBAGHQuantumChemistry2025}, our in-house quantum chemistry software, designed for advanced computational wavefunction-based calculations. It is mainly written in Python, with the bottleneck parts being optimized using Cython and Fortran. It is currently compatible with four interfaces: PySCF\cite{sunLibcintEfficientGeneral2015,sunPySCFPythonbasedSimulations2018,sunRecentDevelopmentsSCF2020}, socutils\cite{wangXubwaSocutils2025}, GAMESS-US\cite{barcaRecentDevelopmentsGeneral2020}, and DIRAC.\cite{bastDIRAC232023} Among other capabilities,
the BAGH software can perform DIP-EOMCC calculations using both NR and various relativistic Hamiltonians. 

The uncontracted dyall.av$n$z ($n=$2,3,4) basis sets\cite{dyallRelativisticDoublezetaTriplezeta2016,dyallRelativisticQuadrupleZetaRevised2006} have been used for the valence DIP energy calculations of inert gas atoms (Ar--Rn) as well as the Cl$_{2}$, Br$_{2}$, HBr and HI molecules.
All diatomic molecules were described using their
experimental bond lengths obtained from Ref.\
\onlinecite{johnsonComputationalChemistryComparison2002}. 
The lowest-energy occupied orbitals corresponding to the 
chemical cores of the elements were frozen in all post-HF
steps of the DIP-EOMCC calculations. The effect of including diffuse functions on the resulting DIPs has been studied by augmenting the dyall.av4z basis set with single, double, and triple sets of diffuse functions. The augmented basis sets are generated using the Dirac software package.\cite{bastDIRAC232023} The calculated  DIP values are extrapolated to the complete basis set (CBS) limit by using the mixed exponential formula of Peterson and Dunning\cite{petersonBenchmarkCalculationsCorrelated1994}, 
\begin{equation}
    E^{x} = E^{\infty} + Ae^{-(x-1)} + Be^{-(x-1)^2}
\end{equation}
where $A$ and $B$ are parameters and $E^{x}$ and $E^{\infty}$ are the total energies for a particular basis ($x$) and at the CBS limit, respectively. 

\begin{figure}[htbp]
   \centering
   \includegraphics[width=0.85\textwidth]{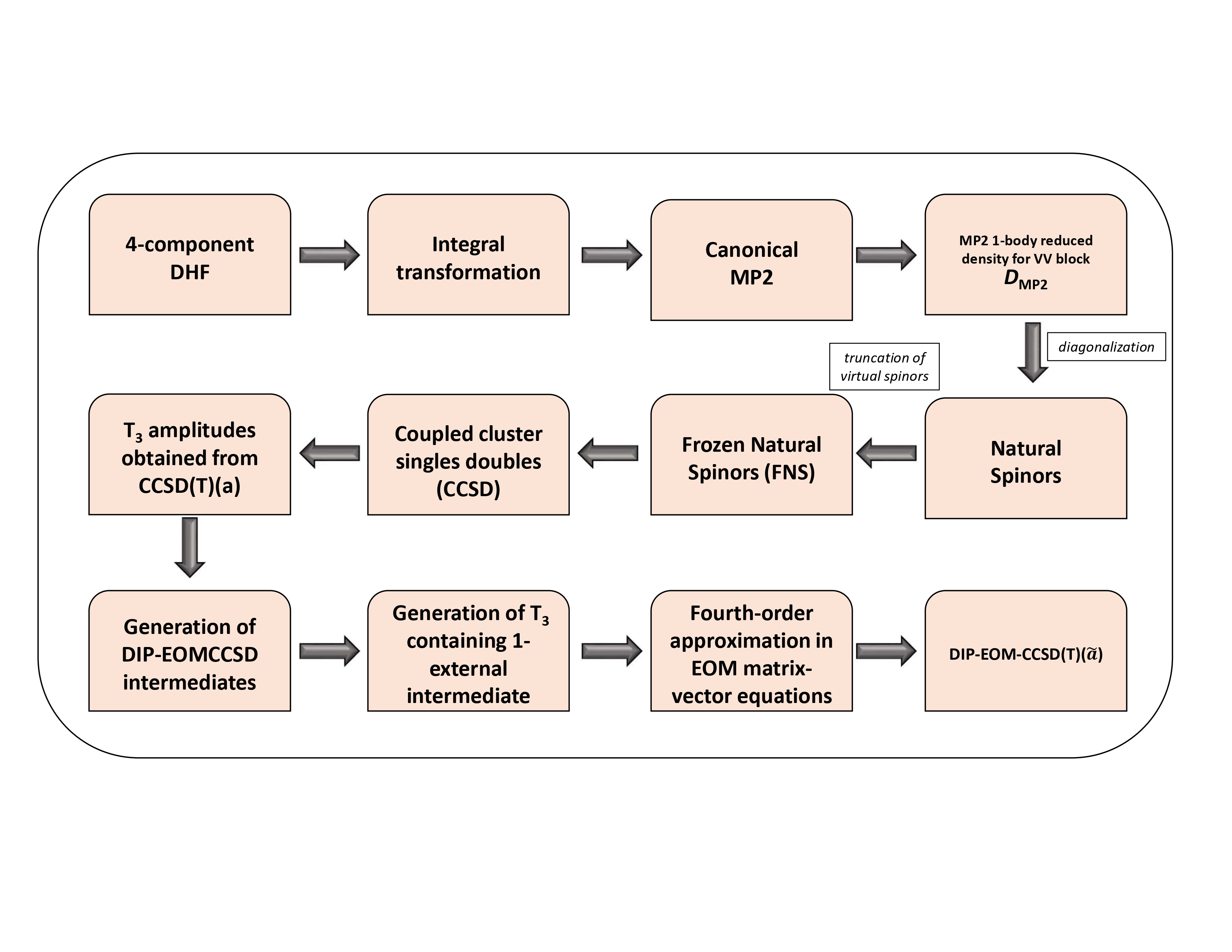}
   \caption{The schematic diagram of the algorithm of the FNS-DIP-EOMCCSD(T)($\tilde{a}$)(4$h$-2$p$) method.}
   \label{fig:algorithm}
\end{figure}

After performing a DHF calculation,
we construct
two-electron integrals of the 
$\langle OO|VV\rangle$ type in the canonical spinor basis in order to perform the MP2 calculation and 
compute the corresponding virtual-virtual block of the
one-body reduced density matrix. The virtual space
is then truncated according to a pre-defined FNS threshold
$n_{\text{thresh}}$ using the recipe described in Section \ref{sec2.3}.
All one- and two-electron integrals are computed and stored
in the truncated natural spinor basis with the help of 
the transformation matrix Eq.\ \ref{eqao2fns}, which allows
us to move from the atomic spinor to FNS space in a computationally
efficient manner. 
The CCSD calculation is performed in the FNS basis, and the amplitudes are noniteratively corrected for $T_{3}$ effects using
Eqs.\ (\ref{eqt1corr})--(\ref{eqt2corr}). In the DIP-EOMCCSD(T)($\tilde{a}$)(4$h$-2$p$) calculations, the $T_{3}$ contributions to the $\bar{h}^{ij}_{am}$ component
of $\bar{H}$ entering Eqs.\ (\ref{eq3h1p}) and (\ref{eq4h2p}) are computed prior to solving the 
DIP-EOMCC eigenvalue problem. A schematic diagram of the algorithm for the FNS-DIP-EOMCCSD(T)($\tilde{a}$)(4$h$-2$p$) approach is provided in FIG. \ref{fig:algorithm}. The FNS-DIP-EOMCCSDT(4$h$-2$p$) and DIP-EOMCCSD(T)(a)(4$h$-2$p$) calculations follow the same algorithm defined in Ref. \onlinecite{dipeomccsdt}, after the one-
and two-electron integrals are generated in FNS basis.
\section{Results and Discussion}
\label{sec4}
\subsection{Choice of FNS threshold}
\label{sec4.1}
\begin{figure}[htbp]
\centering
    \begin{subfigure}{0.45\textwidth} 
        \includegraphics[width=\textwidth]{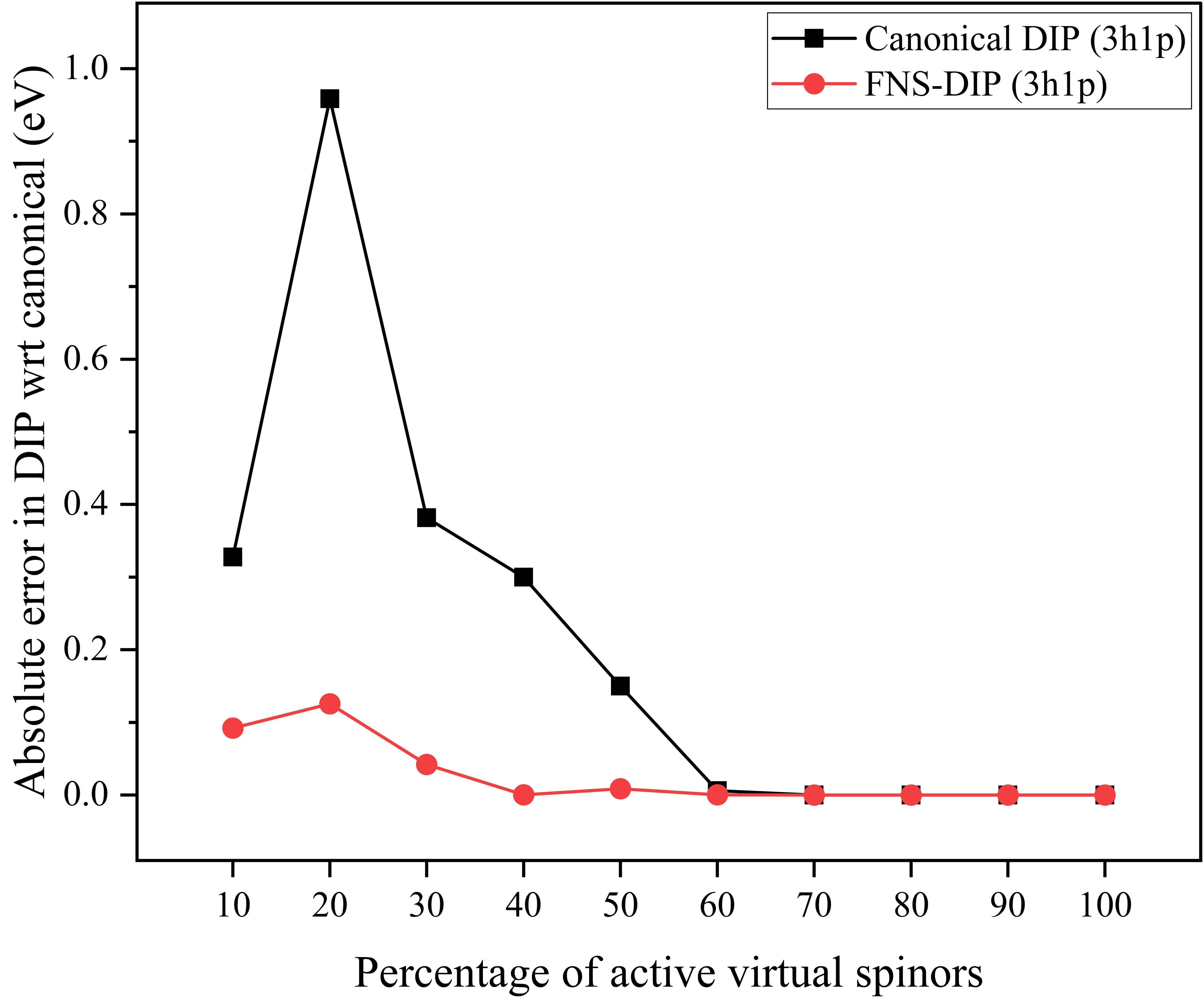}
        \caption{\label{fig:pctnew}}
    \end{subfigure}
    \begin{subfigure}{0.45\textwidth}
        \includegraphics[width=\textwidth]{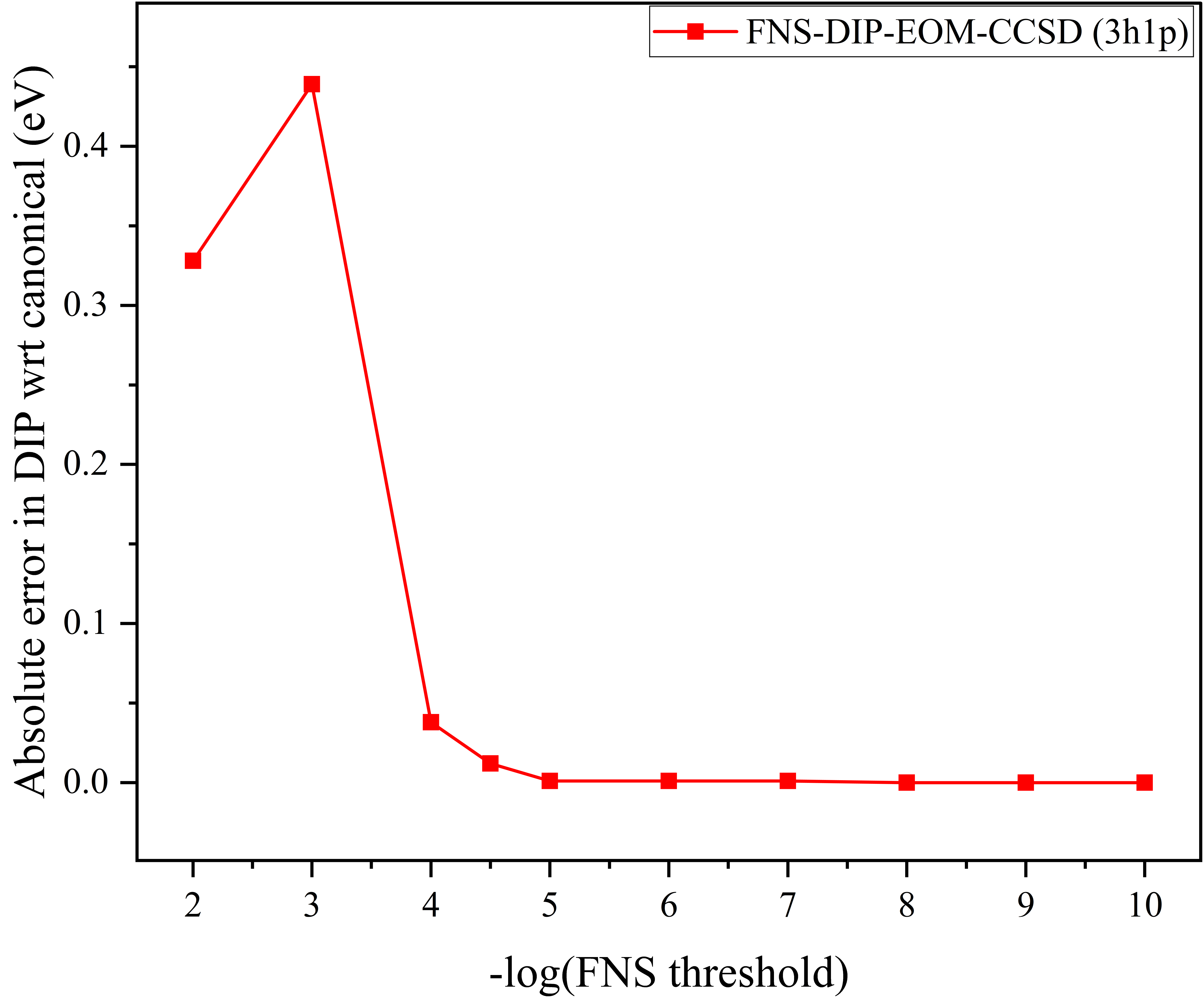}
        \caption{\label{fig:threshnew}}
    \end{subfigure}
    \caption{
    \label{fig:Zn_combined_pct}
    \justifying{The comparison of absolute error of DIP energies (in eV) characterizing the $X\:^{3}\Sigma^{-}$ state
    of \Cltwo{} for the FNS version of DIP-EOMCCSD(3$h$-1$p$) with respect to their respective canonical analogues calculated using dyall.av3z basis set (a) across the percentage of active virtual spinors and (b) across different truncation thresholds.}}
\end{figure}

The main idea behind the FNS approximation is to reduce the size of the virtual space in correlated relativistic calculations. The truncation in the canonical spinor basis leads to a significant loss of accuracy due to the contribution of high-lying virtual spinors to the energy. On the other hand, the virtual spinors in the natural spinor basis are arranged according to their occupancy, which roughly tracks their contribution to the correlation energy. For the case of DIP, where its zeroth-order description (2$h$-TDA) does not involve any virtual orbitals, the FNS basis can give a compact description for the DIP-EOMCCSD(3$h$-1$p$)
and DIP-EOMCCSD(4$h$-2$p$) calculations similar to that observed for the FNS-IP-EOMCCSD method\cite{4c-eomcc-fns}. To illustrate
this point, we present the convergence of the lowest valence
DIP for \Cltwo{}
with respect to the size of the truncated virtual space in 
Fig.\ \ref{fig:pctnew}. It can be seen that the DIP values converge more quickly in the FNS basis and the results approach their canonical counterparts using just 40$\%$ of the total virtual space. In the canonical basis, one needs to include at least 60 $\%$  of the virtual space to achieve a similar level of accuracy. However, the FNS threshold is a better truncation criterion than the size of the virtual space. Fig. \ref{fig:threshnew} presents the convergence of absolute error characterizing
the lowest-lying DIP of \Cltwo{}
computed with DIP-EOMCCSD(3$h$-1$p$)
as a function of the FNS threshold.
The calculation is performed in dyall.av3z basis set and the DIP-EOMCCSD(3$h$-1$p$) values in the untruncated canonical basis have been taken as the reference. From Fig. \ref{fig:threshnew}, the truncation error in the DIP for the $X\:^{3}\Sigma^{-}$ state of \Cltwoplus{}
is less than 0.1 eV using an FNS threshold of $10^{-4}$. With an FNS threshold of $10^{-5}$, the DIP obtained
using the FNS-based
DIP-EOMCCSD(3$h$-1$p$) approach
is virtually identical to its
untruncated counterpart. However, considering the fact that the FNS threshold is directly related to the number of virtual spinors that will be considered for correlation calculations, taking $10^{-5}$ as the FNS threshold for calculations may be a costly choice. The FNS threshold of $10^{-4.5}$ seems to be a good compromise between cost and accuracy
as the
truncation error characterizing the
DIP for the $X\:^{3}\Sigma^{-}$ state of \Cltwoplus{} is as little as 0.01 eV. 
Based on these observations, we adopt
an FNS occupation threshold of $10^{-4.5}$ for all calculations,
unless otherwise mentioned.
\subsection{Basis set convergence}
\label{sec4.2}
To analyze the effect of basis set,
we have compared the DIPs corresponding
to the $X\; {^{3}}\Sigma^{-}$, $a\; {^{1}}\Delta$, $b\; {^{1}}\Sigma^{+}$ and $c\; {^{1}}\Sigma^{-}$ states of \Cltwoplus{}, as obtained with
DIP-EOMCCSD(3$h$-1$p$) using the
dyall.av$n$z ($n=2$,3,4) hierarchy, 
with their experimentally determined
counterparts in Table \ref{tab:cb}.  Furthermore, in order to investigate the effect of diffuse functions on the calculated DIP values, the results obtained
with dyall.av4z have been augmented with single, double, and triple sets of diffuse functions. It can be seen that the DIP values increase with the basis
set cardinality number. For example,
the DIPs characterizing the valence
states of \Cltwoplus{}
grow by roughly 0.3 eV when
the basis set is increased from
dyall.av2z to dyall.av3z. Slighly
smaller changes of $\sim$0.2 eV
are observed when increasing the basis
from dyall.av3z to dyall.av4z.
As shown in Table \ref{tab:cb}, the
DIP values obtained with dyall.av4z
increase by 0.1--0.2 eV when CBS
extrapolations are employed. Thus,
even the dyall.av4z basis results in
significant basis set errors.

In contrast, the inclusion of diffuse
functions in the basis set provides
hardly any effect on the resulting
DIP values.
As shown in Table \ref{tab:cb}, the DIP values computed with DIP-EOMCCSD(3$h$-1$p$) using the dyall.av3z basis show the best agreement with experiment. However, this agreement
vanishes when a larger dyall.av4z basis
is employed. One needs
to use a higher-level treatment of
many-electron correlation effects to obtain
an accurate and systematic behavior of the
DIP values.
\subsection{The effect of the inclusion of higher excitations}
\label{sec4.4}
In order to investigate the effect of the inclusion of higher-rank excitation and
doubly ionizing operators, the DIPs characterizing the $X\:^{3}\Sigma^{-}$, $a\:^{1}\Delta$,
$b\:^{1}\Sigma^{+}$, and $c\:^{1}\Sigma^{-}$
states of \Cltwoplus{} computed with different levels of DIP-EOMCC approximation have been presented in Table \ref{tab:Cl2}. The FNS-DIP-EOMCCSD(3$h$-1$p$) results are calculated in dyall.av3z and dyall.av4z bases. The results obtained from the dyall.av3z basis have been shown for the sake of comparison with the previously reported canonical values by 
Pal and co-workers.\cite{pathakRelativisticDoubleionizationEquationofmotion2020} 
It can be seen that at the FNS-DIP-EOMCCSD(3$h$-1$p$)/dyall.av3z level of theory, the resulting DIPs show good agreement with their experimental counterparts. The FNS-DIP-EOMCCSD(3$h$-1$p$)/dyall.av3z DIP values are also nearly identical to the canonical results reported by Pal and co-workers\cite{pathakRelativisticDoubleionizationEquationofmotion2020} using the same level of theory. However, the DIPs computed
with FNS-DIP-EOMCCSD(3$h$-1$p$)/dyall.av4z are
significantly larger than experiment. 
 The inclusion of the 4$h$-2$p$ excitations
on top of CCSD leads to an underestimation of the DIP values. However, the inclusion of $T_{3}$ correlations within
the DIP-EOMCCSDT(4$h$-2$p$) restores a proper balance between ground and target ($N-2$)-electron state, providing improved agreement
between predicted and experimentally 
determined DIP values. The DIP-EOMCCSD(T)($\tilde{a}$)(4$h$-2$p$) approximation developed in this work, which can be
performed using computational steps that
scale as $\mathscr{N}^{7}$ with the system size, is capable of delivering DIPs for
the relevant states of \Cltwoplus{} in agreement
with their counterparts obtained with the
much more expensive
DIP-EOMCCSDT(4$h$-2$p$) and DIP-EOMCCSD(T)(a)(4$h$-2$p$) approaches.  Therefore, we rely on
the DIP-EOMCCSD(T)($\tilde{a}$)(4$h$-2$p$)/CBS level of theory in order to investigate the 
remaining atomic and molecular systems
of interest in this study. 

\subsection{Benchmarking: atoms and molecules}
\label{sec4.5}
In order to benchmark the performance of DIP-EOMCCSD(T)($\tilde{a}$)(4$h$-2$p$) method, we have calculated the DIPs of the series of inert gas atoms Ar--Rn. The calculated DIP values along with their experimental counterparts are presented in Table \ref{tab:atoms}. Key statistical parameters,
including the maximum absolute deviation (MAD),
mean absolute error (MAE), standard
deviation (STD), and root-mean-squared deviation
(RMSD), are also included in Table \ref{tab:atoms}. Using the dyall.av3z basis, the DIPs computed with DIP-EOMCCSD(3$h$-1$p$) show good agreement with the available experimental results,\cite{NIST_ASD_2024} with
MAD and RMSD values of 0.26 eV and 0.15 eV,
respectively. However, after performing
CBS extrapolations, the error in the DIPs
obtained with DIP-EOMCCSD(3$h$-1$p$) relative
to experiment significantly increase,
resulting in MAD and RMSD values of 0.47 eV
and 0.34 eV, respectively. The increase in error at the DIP-EOMCCSD(3$h$-1$p$) level with the size of the basis is consistent with 
the trend observed for \Cltwo{} discussed in
the previous section.
When the DIP-EOMCCSD(3$h$-1$p$)/CBS method is
replaced by its higher-level 
DIP-EOMCCSD(T)($\tilde{a}$)(4$h$-2$p$)/CBS counterpart, the MAD and RMSD values
relative to experiment for the
DIPs reported in Table \ref{tab:atoms}
reduce to 0.16 eV and 0.06 eV,
respectively.

The DIP-EOMCCSD(T)($\tilde{a}$)(4$h$-2$p$) 
approach is also applied
to obtain the valence DIPs of
Cl$_2$, Br$_2$, HBr, and HI, which have been considered in
previous works.\cite{dipeomccsdt,liRelativisticTwocomponentDouble2025,pathakRelativisticDoubleionizationEquationofmotion2020} 
Similar to atoms, the DIP values at DIP-EOMCCSD(3$h$-1$p$)/dyall.av3z level show good agreement with experiment,\cite{mcconkeyThresholdPhotoelectronsCoincidence1994,fleigTheoreticalExperimentalStudy2008,elandCompleteDoublePhotoionisation2003,yenchaPhotodoubleIonizationHydrogen2004} with a
MAD of 0.35 eV and RMSD of 0.19 eV obtained for
the valence states of \Cltwoplus{}, \Brtwoplus{}, \HBrtwoplus{}, and \HItwoplus{} considered
in Table \ref{tab:molecules}. 
Consistent with previous observations, the
errors obtained with DIP-EOMCCSD(3$h$-1$p$)
increase when the basis set size is increased
and extrapolated toward the CBS limit.
Indeed, the DIPs computed with DIP-EOMCCSD(3$h$-1$p$)/CBS are less accurate relative
to experiment than their counterparts
obtained using the dyall.av3z basis, with 
MAD and RMSD values of 0.66 eV and 0.44 eV,
respectively.
Larger errors are observed, especially for Cl$_2$ and HBr. 
The errors relative to experiment significantly reduce when the
DIP-EOMCCSD(3$h$-1$p$)/CBD method is
replaced by the higher-level DIP-EOMCCSD(T)($\tilde{a}$)(4$h$-2$p$)/CBS approach. 
In particular, when using 
DIP-EOMCCSD(T)($\tilde{a}$)(4$h$-2$p$)/CBS,
the DIPs for \Cltwo{}, \Brtwo{}, HBr, and HI
are characterized by MAD and RMSD values
relative to experiment of 0.30 eV and 0.15 eV,
respectively.
When examining the accuracy of the 
DIP-EOMCCSD(T)($\tilde{a}$)(4$h$-2$p$)/CBS
results reported in Table \ref{tab:molecules}
relative to experiment, we observe slighly
larger errors for molecules than for 
the atoms considered in Table \ref{tab:atoms}.
This difference in performance may be due to
the effects of molecular vibration, which are
not included in the present DIP-EOMCC
calculations.

The Br$_2$ molecule requires special attention as the previous application of the
DIP-EOMCCSDT(4$h$-2$p$) approach based on the one-electron
SFX2C framework (SFX2C1e) resulted in nonnegligible errors.\cite{dipeomccsdt} Table \ref{tab:my_label} presents the DIP values corresponding to the first ten states of \Brtwoplus{}, in which we use the $\Lambda-S$ coupling notation for electronic
states, similar to that used by Fleig et.\ al.\cite{fleigTheoreticalExperimentalStudy2008}
The previous study by Pal and co-workers\cite{pathakRelativisticDoubleionizationEquationofmotion2020} only reported four of these states. Our previous SFX2C1e-based DIP-EOMCCSDT(4$h$-2$p$) study\cite{dipeomccsdt} has not been able to distinguish between the lowest A $0_{g}$ and A $1_{g}$ states due to the neglect of spin-orbit coupling in the SFX2C1e calculations. The DIP-EOMCCSD(T)($\tilde{a}$)(4$h$-2$p$) method based on a 4c-DC Hamiltonian can accurately resolve the lowest A $0_{g}$ and A $1_{g}$ states in agreement with experiment. The next two states, A $2_{g}$ and A $0_{g}$, also show good agreement with the experiment.  As shown by Fleig et. al.\cite{fleigTheoreticalExperimentalStudy2008}, the experimentally observed broad band at 30.3 eV contains contributions from a group of 
electronic states
with energies ranging between 30.1--30.5 eV, 
which complicates the comparison between
individual states and experiment.
However, for all the states considered
in Table \ref{tab:my_label}, the  DIP-EOMCCSD(T)($\tilde{a}$)(4$h$-2$p$) method gives better agreement with experiment than the previously reported DIP-EOMCCSD(3$h$-1$p$) values
of \onlinecite{pathakRelativisticDoubleionizationEquationofmotion2020}. 
It is also worth noting that the 
DIP-EOMCCSD(T)($\tilde{a}$)(4$h$-2$p$)/CBS
results reported in Table \ref{tab:my_label}
are in excellent agreement with the relativistic multireference CI (MRCI) results\cite{fleigTheoreticalExperimentalStudy2008} reported by Fleig et al for all ten 
electronic states of \Brtwoplus{}.
\subsection{The effect of the treatment of relativity}
It is important to investigate the impact
of different levels of treating the relativistic
Hamiltonian on the DIPs obtained in the DIP-EOMCC
calculations.
Table~\ref{tab:atoms_NR_X2C} presents a comparison of DIPs characterizing for Ar, Kr, Xe and Rn atoms using NR, SFX2C1e, and 4c-DC Hamiltonians alongside the experimentally
determined results. All DIPs are obtained
with the DIP-EOMCCSD(T)($\tilde{a}$)(4$h$-2$p$)/CBS calculations. The NR calculations show large errors
relative to experiment, with MAD and RMSD values
of 1.83 eV and 0.57 eV, respectively. 
As one might expect, the largest error is observed for the heaviest atom, Rn. The use of scalar relativity, however, does not improve
the situation. In fact, as shown in Table \ref{tab:atoms_NR_X2C}, the errors relative to experiment characterizing 
the DIPs of Ar--Rn actually increase when
relativistic effects are treated using
the SFX2C1e approach, with MAD and RMSD values
of 2.04 eV and 0.60 eV, respectively.
Based on the results for Ar--Rn reported in
Table \ref{tab:atoms_NR_X2C}, one must use a
more complete 4c-DC Hamiltonian in order to
obtain DIP values that agree with experiment.
The inclusion of the Gaunt and Breit corrections have negligible effect on the DIP values considered in Table \ref{tab:atoms_NR_X2C}.

When considering the \Cltwo{}, \Brtwo{}, HBr,
and HI molecules, we do not observe the same
behavior in DIP values 
as we do for atoms when the treatment of
relativity is improved. 
In particular, the MAD and RMSD values
characterizing the DIPs computed
with DIP-EOMCCSD(T)($\tilde{a}$)(4$h$-2$p$)/CBS
reported in Table \ref{tab:molecules_NR_X2C},
which are 0.30 eV and 0.15 eV, respectively,
are similar to their counterparts
obtained with the SFX2C1e treatments.
As in the case for the Ar--Rn atoms, the inclusion of the Gaunt and Breit
corrections is negligible.
\section{Conclusion}
\label{sec5}
In this work, we have developed and tested a suite of 4c relativistic DIP-EOMCC methods incorporating up to 4$h$–2$p$ excitations and three-body clusters. We have extended the original
DIP-EOMCCSDT(4$h$-2$p$) method and its perturbative DIP-EOMCCSD(T)(a)(4$h$-2$p$) approximation to a fully relativistic regime. 
In addition, we have
introduced a new, low-cost approximation to DIP-EOMCCSD(T)(a)(4$h$-2$p$), abbreviated as DIP-EOMCCSD(T)($\tilde{a}$)(4$h$-2$p$),
which is capable of accurately reproducing the DIPs obtained with
DIP-EOMCCSD(T)(a)(4$h$-2$p$) and DIP-EOMCCSDT(4$h$-2$p$)
using much smaller computational steps that scale as $\mathscr{N}^7$.
This makes the DIP-EOMCCSD(T)($\tilde{a}$)(4$h$-2$p$) approach a
practical tool for 
studying the DIPs of atomic and molecular systems using realistic
basis sets and relativistic Hamiltonians. We have also combined
all of these approaches with the FNS truncation scheme, and we have
showed that this allows us to significantly reduce the computational
cost without compromising accuracy.
Benchmark calculations on inert gas atoms (Ar--Rn) and diatomics (Cl$_2$, Br$_2$, HBr, HI) demonstrate that the proposed DIP-EOMCCSD(T)($\tilde{a}$)(4$h$-2$p$) method produces DIPs in good agreement with experimental results
after CBS extrapolations are performed. It was also found that 
the calculated DIP values are very sensitive to the basis set size,
and in order to compare with experiment, large basis sets or CBS
extrapolations are necessary. The need for large basis sets
highlights the usefulness of the FNS scheme, as we are able to
handle larger calculations in QZ-level basis sets without running into
prohibitive computational or memory bottlenecks.
Furthermore, we have benchmarked the effects of NR and scalar relativistic
treatments against their complete 4c-DC parent, and showed that only
the 4c-DC Hamiltonian is capable of providing robust and accurate
DIPs of atomic and molecular systems, containing heavier elements.
\begin{acknowledgments}
This work has been supported by the SERB-India under
the CRG (Project No.\ CRG/2022/005672) and MATRICS (Project No.\ MTR/2021/000420) schemes awarded to A.K.D.  P.P. acknowledges
support from the Chemical Sciences, Geosciences and Biosciences
Division, Office of Basic Energy Sciences, Office of Science, 
U.S. Department of Energy
(Grant No.\ DE-FG02-01ER15228 to P.P). A.K.D acknowledges the research fellowship funded by the EU NextGenerationEU through the Recovery and Resilience Plan for Slovakia under project No. 09I03-03-V04-00117.
\end{acknowledgments}
\section*{Data Availability}
The data that support the findings of this study are available within the article.

\section*{Supplementary Material}
The DIP values DZ, TZ and QZ basis used for CBS extrapolation and programmable expressions for DIP-EOMCCSD(T)($\tilde{a}$)(4$h$-2$p$) have been provided in the Supplementary Material.
\section*{References}
\bibliographystyle{aipnum4-1}
\renewcommand{\baselinestretch}{1.1}
\bibliography{paper_arxiv}
\newpage
\clearpage
\onecolumngrid
\newpage
\clearpage

\pagebreak
\clearpage
\newpage

\begin{table*}[h]
\caption{
\label{tab:cb}
Basis set convergence of FNS-DIP-EOMCCSD(3$h$-1$p$) DIP values (in eV) of Cl$_2$ molecule in different Dyall basis sets
}
\begin{ruledtabular}
\begin{tabular}{ l c c c c c c c c}
State &dyall.av2z &dyall.av3z &dyall.av4z &s-aug-dyall.av4z &d-aug-dyall.av4z &t-aug-dyall.av4z & CBS\footnotemark[1] &Expt.\cite{mcconkeyThresholdPhotoelectronsCoincidence1994} \\
\hline
$X\; {^{3}}\Sigma^{-}$  &31.03 &31.39 &31.58 &31.59 &31.59 &31.59 & 31.72 &31.13 \\
$a\; {^{1}}\Delta$      &31.58 &31.91 &32.09 &32.09 &32.09 &32.09 & 32.20 &31.74 \\
$b\; {^{1}}\Sigma^{+}$  &31.94 &32.30 &32.48 &32.48 &32.48 &32.48 & 32.70 &32.12 \\
$c\; {^{1}}\Sigma^{-}$  &32.98 &33.32 &33.51 &33.51 &33.51 &33.51 & 33.63 &32.97\\
\end{tabular}
\end{ruledtabular}
\footnotetext[1]{\mbox{The CBS value calculated by extrapolating dyall.av$n$z ($n=2$,3,4) basis sets}}
\end{table*}

\begin{table*}[h]
\caption{
\label{tab:Cl2}
Comparison of DIP values (in eV) of Cl$_2$ molecule in different FNS-DIP-EOMCC methods with preexisting experimental and theoretical results. In the following table,
we adopt at shorthand notation in which the methods
DIP-EOMCCSD(3$h$-1$p$), DIP-EOMCCSD(4$h$-2$p$),
DIP-EOMCCSD(T)($\tilde{a}$)(4$h$-2$p$), DIP-EOMCCSD(T)(a)(4$h$-2$p$),
and DIP-EOMCCSDT(4$h$-2$p$) are abbreviated as
CCSD(3$h$-1$p$), CCSD(4$h$-2$p$), CCSD(T)($\tilde{a}$)(4$h$-2$p$),
CCSD(T)(a)(4$h$-2$p$), and CCSDT(4$h$-2$p$), respectively.
}
\begin{ruledtabular}
\begin{tabular}{ l c c c c c cc cc c }
State &\multicolumn{2}{c}{CCSD(3$h$-1$p$)} &CCSD(3$h$-1$p$)\footnotemark[1]  &CCSD(4$h$-2$p$) &CCSDT(4$h$-2$p$) &{CCSD(T)(a)(4$h$-2$p$)} &{CCSD(T)($\tilde{a}$)(4$h$-2$p$)} & Expt.\cite{mcconkeyThresholdPhotoelectronsCoincidence1994} \\
\cline{2-9}

      &dyall.av3z  &dyall.av4z  & dyall.av3z& dyall.av4z& dyall.av4z&dyall.av4z &dyall.av4z&      \\
\hline
$X\; {^{3}}\Sigma^{-}$ & 31.41 & 31.60 & 31.40 & 30.90 & 31.21 & 31.19 & 31.20 & 31.13 \\
$a\; {^{1}}\Delta$     & 31.91 & 32.09 & 31.91 & 31.41 & 31.71 & 31.70  & 31.72 &  31.74 \\
$b\; {^{1}}\Sigma^{+}$ & 32.41 & 32.59 & 32.29 & 31.79 & 32.09 & 32.07 & 32.10 &  32.12 \\
$c\; {^{1}}\Sigma^{-}$ & 33.32 & 33.51 & 33.32 & 32.86 & 33.14 & 33.13 & 33.13 & 32.97 \\
\end{tabular}
\end{ruledtabular}

\footnotetext[1]{\mbox{Canonical DIP-EOMCCSD(3$h$-1$p$) results taken from Ref. \onlinecite{pathakRelativisticDoubleionizationEquationofmotion2020}, calculated in dyall.av3z basis set.}}

\end{table*}

\begin{table*}[h]
\caption{
\label{tab:atoms}
Errors in DIP energies (in eV) of Ar, Kr, Xe and Rn atoms with respect to the experiment. In the following table,
we adopt at shorthand notation in which 
DIP-EOMCCSD(3$h$-1$p$) and
DIP-EOMCCSD(T)($\tilde{a}$)(4$h$-2$p$) are abbreviated as
CCSD(3$h$-1$p$) and CCSD(T)($\tilde{a}$)(4$h$-2$p$),
respectively.
}
\begin{ruledtabular}
\begin{tabular}{ c l c c c c}
Atom &States &CCSD(3$h$-1$p$) &CCSD(3$h$-1$p$) &CCSD(T)($\tilde{a}$)(4$h$-2$p$) &Expt.\cite{NIST_ASD_2024} \\
       &                       &  dyall.av3z    & CBS     &  CBS    &      \\
\hline
Ar     &${^{3}}P_{2}$          &$-$0.01 &0.40  &$-$0.02 &43.39 \\
       &${^{3}}P_{1}$          &0.00  &0.38  &$-$0.01 &43.53 \\
       &${^{3}}P_{0}$          &0.01  &0.40  &$-$0.01 &43.58 \\
       &${^{1}}D_{2}$          &0.06  &0.41  &0.02  &45.13 \\
       &${^{1}}S_{0}$          &0.14  &0.47  &0.11  &47.51 \\
       &                       &      &      &      &      \\
Kr     &${^{3}}P_{2}$          &$-$0.13 &0.31  &-0.02 &38.36 \\
       &${^{3}}P_{1}$          &$-$0.12 &0.33  &$-$0.02 &38.92 \\
       &${^{3}}P_{0}$          &$-$0.09 &0.35  &0.01  &39.02 \\
       &${^{1}}D_{2}$          &$-$0.05 &0.32  &0.01  &40.18 \\
       &${^{1}}S_{0}$          &$-$0.01 &0.38  &0.07  &42.46 \\
       &                       &      &      &      &      \\
Xe     &${^{3}}P_{2}$          &$-$0.25 &0.25  &0.01  &33.11 \\
       &${^{3}}P_{1}$          &$-$0.26 &0.26  &$-$0.02 &34.32 \\
       &${^{3}}P_{0}$          &$-$0.20 &0.29  &0.01  &34.11 \\
       &${^{1}}D_{2}$          &$-$0.18 &0.27  &0.03  &35.23 \\
       &${^{1}}S_{0}$          &$-$0.13 &0.31  &0.06  &37.58 \\
       &                       &      &      &      &      \\
Rn     &${^{3}}P_{2}$          &$-$0.25 &0.19  &$-$0.16 &29.74 \\
\hline
MAD    &                       &0.26  &0.47  &0.16  &      \\
MAE    &                       &0.12  &0.33  &0.04  &      \\
STD    &                       &0.12  &0.07  &0.06  &      \\
RMSD   &                       &0.15  &0.34  &0.06  &      \\
\end{tabular}
\end{ruledtabular}

\end{table*}

\begin{table*}[h]
\caption{
\label{tab:molecules}
Errors in DIP energies (in eV) of Cl$_2$, Br$_2$. HBr and HI molecules with respect to the experiment.
 In the following table,
we adopt at shorthand notation in which 
DIP-EOMCCSD(3$h$-1$p$) and
DIP-EOMCCSD(T)($\tilde{a}$)(4$h$-2$p$) are abbreviated as
CCSD(3$h$-1$p$) and CCSD(T)($\tilde{a}$)(4$h$-2$p$),
respectively.
}
\begin{ruledtabular}
\begin{tabular}{ c l c c c c }
Molecule &States &CCSD(3$h$-1$p$)  &CCSD(3$h$-1$p$) &CCSD(T)($\tilde{a}$)(4$h$-2$p$) &Expt.\cite{mcconkeyThresholdPhotoelectronsCoincidence1994,fleigTheoreticalExperimentalStudy2008,elandCompleteDoublePhotoionisation2003,yenchaPhotodoubleIonizationHydrogen2004} \\
       &                               &dyall.av3z  &CBS & CBS&  \\
\hline
Cl$_2$ &X\; ${^{3}}\Sigma^{-}$         &0.28  &0.59  &0.21  &31.13 \\
       &a\; ${^{1}}\Delta$             &0.17  &0.46  &0.11  &31.74 \\
       &b\; ${^{1}}\Sigma^{+}$         &0.29  &0.58  &0.13  &32.12 \\
       &c\; ${^{1}}\Sigma^{-}$         &0.35  &0.66  &0.29  &32.97 \\
       &                               &      &      &      &      \\
Br$_2$ &A\;   $0_{g}$                  &$-$0.05 &0.26  &0.02  &28.39 \\
       &A\;   $1_{g}$                  &$-$0.07 &0.26  &$-$0.04 &28.53 \\
       &A\;   $2_{g}$                  &0.07  &0.40  &0.02  &28.91 \\
       &A\;   $0_{g}$                  &$-$0.03 &0.28  &$-$0.02 &29.38 \\
       &                               &      &      &      &      \\
HBr    &X\; ${^{3}}\Sigma^{-}$         &0.17  &0.51  &0.13  &32.62 \\
       &a\; ${^{1}}\Delta$             &0.26  &0.53  &0.24  &33.95 \\
       &b\; ${^{1}}\Sigma^{+}$         &0.30  &0.59  &0.30  &35.19 \\
       &                               &      &      &      &      \\
HI     &X\; ${^{3}}\Sigma_{0}^{-}$     &$-$0.14 &0.26  &0.02  &29.15 \\
       &A\; ${^{3}}\Sigma_{1}^{-}$     &$-$0.14 &0.28  &$-$0.02 &29.37 \\
       &a\; ${^{1}}\Delta$             &$-$0.07 &0.29  &0.07  &30.39 \\
       &b\; ${^{1}}\Sigma^{+}$         &0.01  &0.36  &0.12  &31.64 \\
\hline
MAD    &                               &0.35  &0.66  &0.30  &      \\
MAE    &                               &0.16  &0.42  &0.12  &      \\
STD    &                               &0.17  &0.15  &0.11  &      \\
RMSD   &                               &0.19  &0.44  &0.15  &      \\
\end{tabular}
\end{ruledtabular}

\end{table*}

\begin{table*}[h]
    \caption{The comparison of DIP-EOMCCSD(T)($\tilde{a}$)(4$h$-2$p$)/CBS values of Br$_2$ with experiment and previous theoretical results}
    \label{tab:my_label}
    \centering
    \begin{ruledtabular}
    \begin{tabular}{ccccc}
       States  & CCSD(3$h$-1$p$)\footnotemark[1]&CCSD(T)($\tilde{a}$)(4$h$-2$p$)  & MRCI\footnotemark[2] & Expt.\footnotemark[2]\\
       \hline
       
        A $0_{g}$ &       &28.41  &28.39  &28.39  \\
        A $1_{g}$ &28.47  &28.49  &28.54  &28.53  \\
        A $2_{g}$ &29.04  &28.93  &29.01  &28.91  \\
        A $0_{g}$ &29.52  &29.36  &29.45  &29.38  \\
        B $0_{u}$ &       &29.77  &29.78  &       \\
        B $3_{u}$ &       &29.80  &29.81  &       \\
        B $2_{u}$ &29.79  &30.16  &30.16  &30.30  \\
        B $1_{u}$ &       &30.24  &30.24  &       \\
        B $0_{u}$ &       &30.47  &30.50  &       \\
        B $1_{u}$ &       &30.51  &30.52  &       \\
    \end{tabular}

    \end{ruledtabular}
    \footnotetext[1]{\mbox{ Taken from Ref. \onlinecite{pathakRelativisticDoubleionizationEquationofmotion2020}}}
    \footnotetext[2]{\mbox{ Taken from Ref. \onlinecite{fleigTheoreticalExperimentalStudy2008}}}
\end{table*}

\begin{table*}[h]
\caption{
\label{tab:atoms_NR_X2C}
Comparison of DIP energies (in eV) of Ar, Kr, Xe, and Rn atoms using DIP-EOMCCSD(T)($\tilde{a}$)(4$h$-2$p$) at CBS level with NR and various relativistic Hamiltonians.
}
\begin{ruledtabular}
\begin{tabular}{ c l c c c c c c}
Atom    &States &NR(CBS) &SFX2C1e(CBS) &4c-DC(CBS) &4c-DCG(CBS) & 4c-DCB(CBS) &Expt.\cite{NIST_ASD_2024} \\
\hline
Ar     &${^{3}}P_{2}$           &43.47  &43.44 &43.37 &43.37 &43.37 &43.39 \\
       &${^{3}}P_{1}$           &43.47  &43.44 &43.51 &43.49 &43.51 &43.53 \\
       &${^{3}}P_{0}$           &43.47  &43.44 &43.58 &43.56 &43.56 &43.58 \\
       &${^{1}}D_{2}$           &45.17  &45.14 &45.15 &45.13 &45.15 &45.13 \\
       &${^{1}}S_{0}$           &47.63  &47.61 &47.62 &47.62 &47.62 &47.51 \\
       &                        &       &      &      &      &      &      \\
Kr     &${^{3}}P_{2}$           &38.65  &38.64 &38.34 &38.33 &38.33 &38.36 \\
       &${^{3}}P_{1}$           &38.65  &38.64 &38.90 &38.88 &38.88 &38.92 \\
       &${^{3}}P_{0}$           &38.65  &38.64 &39.03 &39.00 &39.02 &39.02 \\
       &${^{1}}D_{2}$           &40.11  &40.11 &40.19 &40.16 &40.15 &40.18 \\
       &${^{1}}S_{0}$           &42.34  &42.38 &42.53 &42.51 &42.51 &42.46 \\
       &                        &       &      &      &      &      &      \\
Xe     &${^{3}}P_{2}$           &33.74  &33.78 &33.11 &33.11 &33.10 &33.11 \\
       &${^{3}}P_{1}$           &33.74  &33.78 &34.30 &34.26 &34.28 &34.32 \\
       &${^{3}}P_{0}$           &33.74  &33.78 &34.13 &34.11 &34.11 &34.11 \\
       &${^{1}}D_{2}$           &34.91  &34.97 &35.26 &35.23 &35.23 &35.23 \\
       &${^{1}}S_{0}$           &36.81  &36.95 &37.64 &37.59 &37.61 &37.58 \\
       &                        &       &      &      &      &      &      \\
Rn     &${^{3}}P_{2}$           &31.57  &31.78 &29.58 &29.56 &29.57 &29.74 \\
\hline
MAD    &                        &1.83   &2.04  &0.16  &0.18  &0.17  &      \\
MAE    &                        &0.38   &0.37  &0.04  &0.04  &0.04  &      \\
STD    &                        &0.59   &0.62  &0.06  &0.06  &0.06  &      \\
RMSD   &                        &0.57   &0.60  &0.06  &0.06  &0.06  &      \\

\end{tabular} 
\end{ruledtabular}
\end{table*}

\begin{table*}[h]
\caption{
\label{tab:molecules_NR_X2C}
Comparison of DIP energies (in eV) of Cl$_2$, Br$_2$. HBr and HI molecules using DIP-EOMCCSD(T)($\tilde{a}$)(4$h$-2$p$) at CBS level with NR and various relativistic Hamiltonians.
}
\begin{ruledtabular}
\begin{tabular}{ c l c c c c c c}
Molecule &States &NR(CBS) &SFX2C1e(CBS) &4c(CBS) &DCG(CBS) &DCB(CBS) &Expt.\cite{mcconkeyThresholdPhotoelectronsCoincidence1994,fleigTheoreticalExperimentalStudy2008,elandCompleteDoublePhotoionisation2003,yenchaPhotodoubleIonizationHydrogen2004} \\
\hline
Cl$_2$ &X\; ${^{3}}\Sigma^{-}$       &31.39 &31.31 &31.34 &31.32 &31.32 &31.13 \\
       &a\; ${^{1}}\Delta$           &31.87 &31.83 &31.85 &31.84 &31.83 &31.74 \\
       &b\; ${^{1}}\Sigma^{+}$       &32.26 &32.22 &32.25 &32.23 &32.23 &32.12 \\
       &c\; ${^{1}}\Sigma^{-}$       &33.32 &33.28 &33.26 &33.24 &33.24 &32.97 \\
       &                             &      &      &      &      &      &      \\
Br$_2$ &A\;   $0_{g}$                &28.56 &28.51 &28.41 &28.40 &28.40 &28.39 \\
       &A\;   $1_{g}$                &28.56 &28.51 &28.49 &28.48 &28.48 &28.53 \\
       &A\;   $2_{g}$                &29.00 &28.95 &28.93 &28.92 &28.91 &28.91 \\
       &A\;   $0_{g}$                &29.34 &29.29 &29.36 &29.38 &29.38 &29.38 \\
       &                             &      &      &      &      &      &      \\
HBr    &X\; ${^{3}}\Sigma^{-}$       &32.89 &32.85 &32.75 &32.74 &32.75 &32.62 \\
       &a\; ${^{1}}\Delta$           &34.23 &34.21 &34.19 &34.18 &34.17 &33.95 \\
       &b\; ${^{1}}\Sigma^{+}$       &35.49 &35.48 &35.49 &35.48 &35.49 &35.19 \\
       &                             &      &      &      &      &      &      \\
HI     &X\; ${^{3}}\Sigma_{0}^{-}$   &29.40 &29.39 &29.17 &29.16 &29.18 &29.15 \\
       &A\; ${^{3}}\Sigma_{1}^{-}$   &29.40 &29.39 &29.35 &29.33 &29.34 &29.37 \\
       &a\; ${^{1}}\Delta$           &30.50 &30.50 &30.46 &30.45 &30.44 &30.39 \\
       &b\; ${^{1}}\Sigma^{+}$       &31.58 &31.58 &31.76 &31.74 &31.75 &31.64 \\
\hline
MAD    &                             &0.34  &0.30  &0.30  &0.29  &0.30  &      \\
MAE    &                             &0.17  &0.14  &0.12  &0.11  &0.11  &      \\
STD    &                             &0.16  &0.12  &0.11  &0.11  &0.11  &      \\
RMSD   &                             &0.20  &0.17  &0.15  &0.14  &0.14  &      \\
\end{tabular}
\end{ruledtabular}
\end{table*}

\end{document}